# Transmit Power Optimization in Optical Coherent Transmission Systems: Analytical, Simulation, and Experimental Results

Ramin Hashemi
Department of Electrical Engineering
Amirkabir University of Technology
(Tehran Polytechnic)
Tehran, Iran
raminhashemi@aut.ac.ir

Mehdi Habibi
Department of Electrical Engineering
Amirkabir University of Technology
(Tehran Polytechnic)
Tehran, Iran
mehdi_habibi@aut.ac.ir

Hamzeh Beyranvand[*]
Department of Electrical Engineering
Amirkabir University of Technology
(Tehran Polytechnic)
Tehran, Iran
beyranvand@aut.ac.ir

Ali Emami, Mahdi Hashemi and Davood Ranjbar Rafie
Iran Telecommunication Research Center
Tehran, Iran
emami@itrc.ac.ir, mhashemi@itrc.ac.ir, ranjbar@itrc.ac.ir

*Abstract—* In this paper, we propose to use discretized version of the so-called Enhanced Gaussian Noise (EGN) model to estimate the non-linearity effects of fiber on the performance of optical coherent and uncompensated transmission (CUT) systems. By computing the power of non-linear interference noise and considering optical amplifier noise, we obtain the signal-to-noise (SNR) ratio and achievable rate of CUT. To allocate power of each CUT channel, we consider two optimization problems with the objectives of maximizing minimum SNR margin and achievable rate. We show that by using the discretized EGN model, the complexity of the introduced optimization problems is reduced compared with the existing optimization problems developed based on the so-called discretized Gaussian Noise (GN) model. In addition, the optimization based on the disctretized EGN model leads to a better SNR and achievable rate. We validate our analytical results with simulations and experimental results. We simulate a five-channel coherent system on OptiSystem software, where a close agreement is observed between optimizations and simulations. Furthermore, we measured SNR of commercial 100Gbps coherent transmitter over 300 km single-mode fiber (SMF) and non-zero dispersion shifted fiber (NZDSF), by considering single-channel and three-channel coherent systems. We observe there are performance gaps between experimental and analytical results, which is mainly due to other sources of noise such as transmitter imperfection noise, thermal noise, and shot noise, in experiments. By including these sources of noise in the analytical model, the gaps between analytical and experimental results are reduced.

*Optical Coherent Transmission Systems, Fiber Non-linear Interference Noise, Power Allocation, Minimum Signal-to-Noise-Ratio (SNR) Margin, Maximum Achievable Capacity.*

---

[*] Corresponding Author

## I. INTRODUCTION

Recently the emerging Coherent and Uncompensated Transmission (CUT) systems in optical fibers have attracted much attention [1]. In addition, bandwidth variable transponders (BVTs) and bandwidth variable wavelength selective switch (BV-WSS) are other two key components to realize Elastic Optical Networks (EONs) based on CUT [2]. Contrary to the conventional Dense Wavelength Division Multiplexing (DWDM) networks in which fiber bandwidth is divided into fixed-grid size channels, in EONs bandwidth is divided into fine-grained slices, and channel bandwidth is determined based on the transmission rate of client. In fact, EONs bring higher flexibility and adaptability to long-haul transmission systems, by selecting channel bandwidth, modulation format, the type of error correction coding, and routing based on the network traffic load and physical layer status [3]-[5]. It has been shown that modeling the effect of fiber impairments on the performance of CUT systems used in EONs will increase the efficiency of resource allocation algorithms [6].

Recently, advanced Digital Signal Processing (DSP) techniques are being utilized at the receiver (Rx) of CUT systems to compensate total dispersion accumulated during the transmission. Consequently, there is no need to use expensive Dispersion Compensating Fibers (DCF). In CUT, the interaction among the non-linearity effects and accumulated dispersion of fiber leads to higher transmission reach [7]. This is due to the fact that fiber dispersion mitigates the non-linearity impairments by reducing the coherency of non-linear interferences [8]. In addition, fiber non-linearity impairments can be mitigated at the Rx of CUT by using advanced DSP techniques.

Estimating the transmission reach of CUT is of paramount importance in the design of EONs and DWDM networks. To this aim, Signal-to-Noise Ratio (SNR) of received signal must be calculated at the end of optical path (lightpath), which must be greater than the minimum required SNR ($SNR_{th}$) of the used modulation format. However, estimating the SNR of CUT system is not straightforward and all its linear and nonlinear noise sources must be modeled.

Recently, in [8], [9] an analytical model has been introduced to estimate the power spectral density of noise generated by the fiber non-linearity effects, which is referred to as Gaussian Noise (GN) model. In this model, self-phase modulation (SPM), cross-phase modulation (XPM), and four-wave mixing (FWM) effects are considered as non-linear noise. The SNR of CUT at the end of lightpath can be computed based on GN model by considering the Amplified Spontaneous Emission (ASE) noise generated by Erbium-Doped Fiber Amplifiers (EDFA) and non-linear noise of fiber, which is referred to as Non-Linear Interference Noise (NLIN).

It has been shown that the GN model overestimates NLIN and does not depend on the modulation format of CUT system [10]. In [11], authors introduced some correction terms to tackle with the overestimation problem of GN model and proposed a number of parameters which depend on the modulation format of CUT system. The model presented in [11] is called Enhanced GN (EGN) model, and authors showed by extensive simulations that EGN can be used to estimate NLIN with an acceptable accuracy.

In [12], it has been shown that optimizing transmission power of CUT systems leads to higher achievable rate and SNR (or equivalently longer transmission reach). The authors used the discretized GN model to optimize the power of CUT DWDM system by proposing two optimization problems, with the objectives of maximizing the total achievable rate or the minimum SNR margin (i.e., the gap between SNR and $SNR_{th}$).

In this paper, we optimize the transmission power of CUT DWDM system by using EGN model instead of GN model. We will show that by using EGN model, higher SNR and achievable rate are obtained in CUT DWDM system. In addition, we propose to use discretized EGN model and remove its multi-channel interference terms, which reduces the complexity of the optimization problem. It is worth mentioning that based on the simulation results reported in [11] the accuracy of SNR estimation is not degraded by ignoring the multi-channel interference terms.

We compare our analytical results with simulation and experimental results. We observe that analytical and simulation results are close while there is a gap between experiment and analytical results. In order to reduce the performance gap between experimental and analytical results, we consider the transmitter imperfection and other sources of noise, which in turns decreases the gap between analytical model and experimental results.

The rest of this paper is organized as follows. In Section II, we review EGN model and present its discretized version, which is used to estimate SNR of CUT systems. In Section III, we present the EGN based power optimization problems to maximize minimum SNR margin and total achievable rate of CUT DWDM systems. In section V, the details of experimental setups are explained. Numerical results are presented in Section V. Finally, the paper is concluded in Section VI.

## II. DISCRETE EGN MODEL

It has been shown that nonlinear noise in fibers can be expressed as a sum of three types of interferences:

• Self-channel interference (SCI): It is nonlinear interference (NLI) caused by four-wave mixing (FWM) occurring among frequency components of channel-of-interest (COI).

• Cross-channel interference (XCI): It is NLI caused by FWM occurring among frequency components of an interferer channel (INT) and COI.

• Multi-channel interference (MCI): It is NLI caused by FWM occurring among frequency components of three INTs, or two INTs and COI.

In EGN-model, each type of interference can be expressed as a sum of two terms, a main term which represents the interference obtained by GN-model and a correction term to remove the Gaussianity assumption of the signal and taking into account the 4th and 6th moments of transmitted symbols, which depend on the modulation format.

In most cases, MCI has negligible value and considering only SCI and XCI gives almost the acceptable value of NLI. Assuming spectral granularity $\Delta f$ and symbol rate R for channels, SCI of COI with center frequency $f_{COI}$ can be written as [3]:

$$G_{SCI}^{EGN} = P_{COI}^3 \left[ \kappa_1(f) + \Phi_{COI}\kappa_2(f) + \Psi_{COI}\kappa_3(f) \right], \quad (1)$$

$$\kappa_1(f) = \frac{16}{27}R^3 \int_{b_1}^{b_2}\int_{b_1}^{b_2} |\mu(f_1,f_2,f)|^2 |s_{COI}(f_1)|^2 \\ |s_{COI}(f_2)|^2 |s_{COI}(f_1+f_2-f)|^2 \, df_1 df_2, \quad (2)$$

$$\kappa_2(f) = \frac{80}{81}R^2 \int_{b_1}^{b_2}\int_{b_1}^{b_2}\int_{b_1}^{b_2} |s_{COI}(f_1)|^2 \, s_{COI}(f_2) \\ s_{COI}^*(f_2') s_{COI}^*(f_1+f_2-f) \, s_{COI}(f_1+f_2'-f) \\ \mu(f_1,f_2,f)\mu^*(f_1,f_2',f) df_1 df_2 df_2' \\ + \frac{16}{27}R^2 \int_{b_1}^{b_2}\int_{b_1}^{b_2}\int_{b_1}^{b_2} |s_{COI}(f_1+f_2-f)|^2 \, s_{COI}(f_1) \\ s_{COI}(f_2) s_{COI}^*(f_1+f_2-f_2') \, s_{COI}^*(f_2') \\ \mu^*(f_1+f_2-f_2',f_2',f)\mu(f_1,f_2,f) df_1 df_2 df_2', \quad (3)$$

$$\kappa_3(f) = \frac{16}{81}R \int_{b_1}^{b_2}\int_{b_1}^{b_2}\int_{b_1}^{b_2}\int_{b_1}^{b_2} s_{COI}(f_1) s_{COI}(f_2) \, s_{COI}^*(f_1+f_2-f) \\ s_{COI}^*(f_1') s_{COI}^*(f_2') \, s_{COI}(f_1'+f_2'-f) \\ \mu^*(f_1',f_2',f)\mu(f_1,f_2,f) df_1 df_2 df_1' df_2'. \quad (4)$$

Assuming lumped amplification and identical spans, $\mu(f_1,f_2,f)$ is given by [3]

$$\mu(f_1,f_2,f) = \gamma \frac{1-e^{-2\alpha L_s} e^{j4\pi^2\beta_2(f_1-f)(f_2-f)L_s}}{2\alpha - j4\pi^2\beta_2(f_1-f)(f_2-f)} \times \\ \frac{\sin\left(2\pi^2\beta_2(f_1-f)(f_2-f)N_s L_s\right)}{\sin\left(2\pi^2\beta_2(f_1-f)(f_2-f)L_s\right)} \times \\ e^{j2\pi^2\beta_2(f_1-f)(f_2-f)(N_s-1)L_s}. \quad (5)$$

In (1)–(5), the integral bounds are $b_1 = f_{COI} - R/2$ and $b_2 = f_{COI} + R/2$. $s_{COI}(f)$ is Fourier transform of the pulse used by COI, and we assumed to be rectangular with bandwidth $R$ and top value $1/R$. $P_{COI}$ is power of COI, $\Phi_{COI}$ and $\Psi_{COI}$ are parameters related to the modulation assigned to COI given in [11], $\gamma$ is fiber non-linearity coefficient, $\alpha$ is optical field fiber loss, $\beta_2$ is second order fiber distortion, $L_s$ is span length, and $N_s$ is the number of spans.

It is worth mentioning that $\kappa_1(f)$ is the main term of SCI, and $\kappa_2(f)$ and $\kappa_3(f)$ are the correction terms. Furthermore, XCI in EGN-model can be written as [11]

$$G_{XCI}^{EGN} = P_{COI}P_{INT}^2 \left[\kappa_{11}(f)+\Phi_{INT}\kappa_{12}(f)\right]+ \\ P_{COI}^2 P_{INT}\left[\kappa_{21}(f)+\Phi_{COI}\kappa_{22}(f)\right]+ \\ P_{COI}^2 P_{INT}\left[\kappa_{31}(f)+\Phi_{COI}\kappa_{32}(f)\right]+ \\ P_{INT}^3\left[\kappa_{41}(f)+\Phi_{INT}\kappa_{42}(f)+\Psi_{INT}\kappa_{43}(f)\right], \quad (6)$$

where $P_{INT}$ is power of interfering channel (INT), and $\Phi_{INT}$ and $\Psi_{INT}$ are parameters related to modulation assigned to INT. In XCI, $\kappa_{m1}$ for $m=1,..,4$ are the main terms and the rest are the correction terms. The complete set of equations for XCI and values of $\Phi$ and $\Psi$ for different modulations can be found in [11]. The main reason behind the definition of discretized EGN model is that by using this model we eventually have a look-up table in terms of channel index. As a consequence, the NLI of each channel is pre-computed in vectors and matrices, instead of computing integrations of EGN model.

By ignoring MCI terms, the total NLIN on COI is given by

$$P_c^{NL} = P_c^3 D_1(c) + \sum_{n=1,n\neq c}^{N}[P_cP_n^2 D_2(c,n)+P_c^2 P_n D_3(c,n)+ \\ P_n^3 D_4(c,n)] \quad (7)$$

where $P_c$ and $P_n$ denote the power of channels $c$ and n, respectively, and $D_1(c)$ is the SCI noise in $c$ th channel, which is obtained as follows

$$D_1(c) = \int_{b_{c1}}^{b_{c2}} \kappa_1(f) + \Phi_c\kappa_2(f) + \Psi_c\kappa_3(f) df. \quad (8)$$

where $b_{c1} = f_0 + \Delta f/2 + (c-1)\Delta f$ and $b_{c2} = f_0 + \Delta f/2 + c.\Delta f$.

In addition, $D_2(c,n)$, $D_3(c,n)$, and $D_4(c,n)$ are the XCI terms generated by $n$ th channel on $c$ th. These terms are obtained as follows

$$D_2(c,n) = \int_{b_{c1}}^{b_{c2}} \kappa_{11}(f) + \Phi_n\kappa_{12}(f) df \quad (9)$$

$$D_3(c,n) = \int_{b_{c1}}^{b_{c2}} \kappa_{21}(f) + \Phi_c\kappa_{22}(f) df + \\ \int_{b_{c1}}^{b_{c2}} \kappa_{31}(f) + \Phi_c\kappa_{32}(f) df, \quad (10)$$

$$D_4(c,n) = \int_{b_{c1}}^{b_{c2}} \kappa_{41}(f) + \Phi_n\kappa_{42}(f) + \Psi_n\kappa_{43}(f) df. \quad (11)$$

In (7)–(11), index $c$ represents COI with $f_{COI} = f_0 + \Delta f/2 + (c-1)\Delta f$, where $f_0$ is the lowest available frequency of the fiber and index $n$ represents INT with $f_{INT} = f_0 + \Delta f /2 + (n-1)\Delta f$.

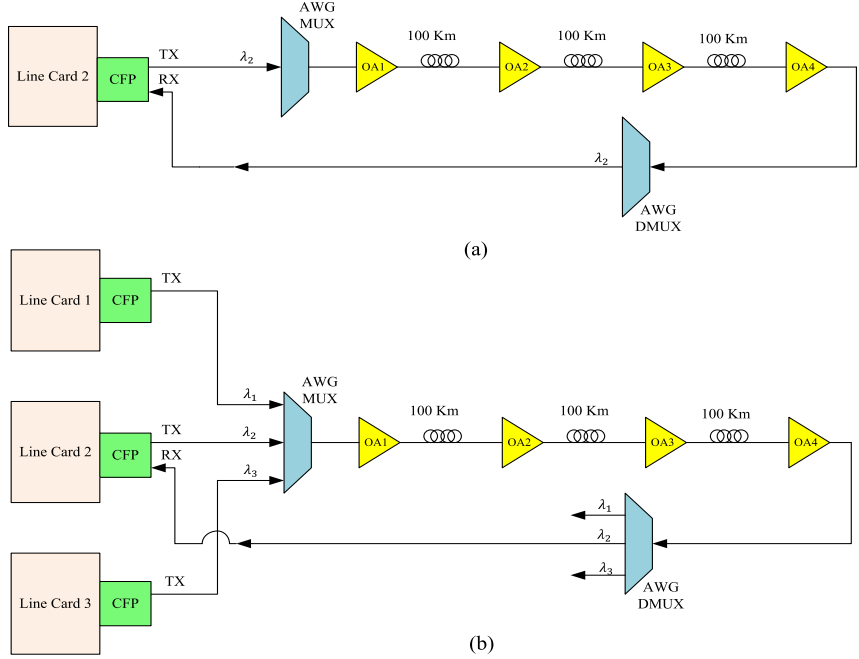

Figure 1. Experimental setup, (a) single channel and (b) three-channel scenarios

In order to obtain SNR, in addition to NLIN power, the total ASE noise generated by optical amplifiers must be computed. The ASE noise of EDFA is given by [13]

$$P_c^{ASE} = R_c h f \left(e^{2\alpha L_s} - 1\right) N_s F, \quad (12)$$

where $h$ is the Planck's constant, $R_c$ is baud-rate of COI, and $F$ is the noise figure of EDFA.

By taking into account both ASE and NLIN, the SNR of COI at the receiver is obtained as follows

$$SNR_c = \frac{P_c}{P_c^{ASE} + P_c^{NL}}. \quad (13)$$

In the next section, we present two optimization problems to allocate power of each DWDM channel by using the above SNR relation which is obtained based on the presented discretized EGN model. Henceforth, we refer to this model as DC-EGN model. It is worth to emphasize that although in the DC-EGN model we have ignored MCI terms, however, as shown in [11] MCI terms in DWDM systems are negligible.

### III. POWER OPTIMIZATION IN DWDM SYSTEMS USING DC-EGN MODEL

*A. Maximizing Minimum SNR Margin*

In this section, we present a power optimization problem based on the introduced DC-EGN model with the objective of maximizing the minimum achievable capacity among channels. Here, we reformulate (13) for $c$ th DWDM channel as a function of $\mathbf{x}$ as follows:

$$SNR_c(\mathbf{x}) = \frac{x_c}{P_c^{ASE} + P_c^{NL}(\mathbf{x})} \quad (14)$$

where $x$ denotes the transmission power in which $\mathbf{x} = [x_1...x_N]^T \in \Re_+^N$ where $N$ is the number of channels, and $P_c^{NL}(\mathbf{x})$ is the nonlinear interference in $c$ th channel which is estimated based on the DC-EGN model given by:

$$P_c^{NL}(\mathbf{x}) = x_c^3 D_1(c) + \sum_{\substack{n=1\\n\neq c}}^{N}[x_c x_n^2 D_2(c,n) + x_c^2 x_n D_3(c,n) + x_n^3 D_4(c,n)] \quad (15)$$

The SNR margin of $c$ th channel is defined as follows [12]:

$$M_c(\mathbf{x}) = \frac{SNR_c(\mathbf{x})}{SNR_{req,c}} \quad (16)$$

where $SNR_{req,c}$ is the required SNR of $c$ th channel for a specific modulation format and coding scheme. The min-max problem which is equivalent to maximization of minimum SNR margin among channels will be:

$$P1.1: \min_{x} \max_{c \in \mathbf{N}} M_c(\mathbf{x})^{-1} \quad (17)$$

where $\mathbf{N} = \{1,2,...N\}$ denotes the set of the channels. The above problem is non-convex and non-linear. To deal with this problem we inspire the same approach proposed in [14]. By using this approach, we apply the variable change $\mathbf{x} = exp(\mathbf{y}), \mathbf{y} \in \Re_+^N$. Thus, the problem P1.1 is modified as follows

$$P1.2: \min_{\mathbf{y}} \max_{c \in \mathbf{N}} \frac{SNR_{req,c}\left(P_c^{ASE} + P_c^{NL}(e^{\mathbf{y}})\right)}{e^{y_n}} \quad (18)$$

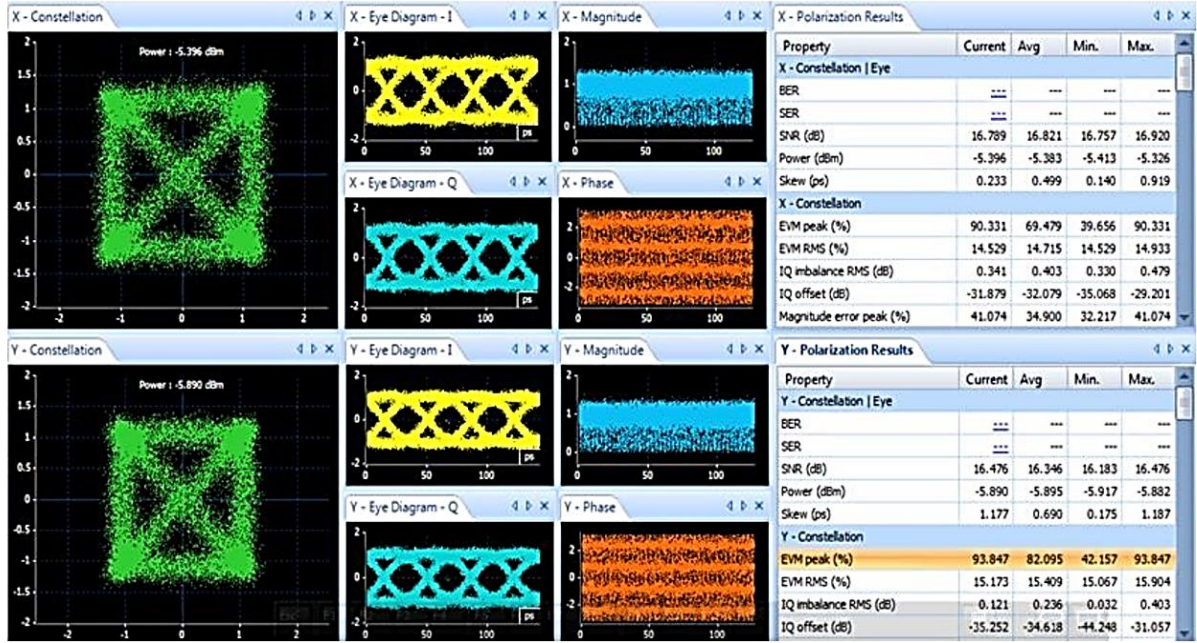

Figure 2. Characteristics of transmitted signals of CFP captured by OMA

where $P_c^{NL}(e^{\mathbf{y}})$ is:

$$P_c^{NL}(e^{\mathbf{y}}) = e^{3y_c} D_1(c) + \sum_{n=1, n \neq c}^{N} [e^{y_c + 2y_n} D_2(c,n) + e^{2y_c + y_n} D_3(c,n) + e^{3y_n} D_4(c,n)] \quad (19)$$

Problem P1.2 is still non-convex. Because $\log(z)$ is a monotonic function, then minimizing $z$ is equivalent to minimizing $\log(z)$, thus P1.2 is equivalent to:

$$P1.3: \min_{y} \max_{c \in \mathbf{N}} \; \log(SNR_{req,c}) + \log((P_c^{ASE} + P_c^{NL}(e^{\mathbf{y}})) - y_c \quad (20)$$

As expressed in [14], by introducing slack variable $s$, P1.3 is equivalent to the following convex problem:

$$P2: \min_{\mathbf{y},s} s \quad (21)$$

s.t. $\log(SNR_{req,c}) + \log((P_c^{ASE} + P_c^{NL}(e^{\mathbf{y}})) - y_c - s \leq 0 \quad (22)$

The inequality constraints of the above problem may be in a region where the objective function becomes infinite which is not desirable. To tackle this issue a barrier function is used and accordingly the following twice differentiable convex problem is obtained:

$$P3: \min_{\mathbf{y},s} s - \frac{1}{t} \sum_{c \in \mathbf{N}} [\log\log(SNR_{req,c}) - \log((P_c^{ASE} + P_c^{NL}(e^{\mathbf{y}}))) + y_c + s] \quad (23)$$

for $t \to \infty$. As problems P2 and P3 are convex optimization, we can solve them easily by using CVX, a well-known optimization tool [14].

*B. Maximizing Achievable Rate*

The total achievable rate of all DWDM channels are given by:

$$C(\mathbf{y}) = \sum_{c \in \mathbf{N}} \log\left(1 + \frac{e^{y_c}}{P_c^{ASE} + P_c^{NL}(e^{\mathbf{y}})}\right) \quad (24)$$

For high SNR scenarios, we have $\log(1 + SNR) \approx \log(SNR)$ thus, the problem of optimizing sum rate at higher SNRs is given by:

$$P4: \max_{\mathbf{y}} \sum_{c \in \mathbf{N}} \left(y_c - \log\left(P_c^{ASE} + P_c^{NL}(e^{\mathbf{y}})\right)\right). \quad (25)$$

which can be solved in the same approach presented in the former subsection.

## IV. EXPERIMENTAL SETUP

We have considered two arrangements to measure SNR of commercial optical coherent transceivers over three spans of 100 km fibers, and two fiber types (SMF and NZDSF) were used for each arrangement. In the first arrangement, we have tested a single channel 100 Gbps coherent transmission system, and in the second one, we investigated a three-channel coherent transmission system.

In Figure 1 schematic of the experimental setups are shown. In these schematics, line card means an electronic board encapsulating the client data with rate of 100 Gbps and Ethernet framing into OTN framing based on the ITU-T G.709 standard [15]. It is worth noting that these line cards are known as transponders, in which pluggable module, C Form-factor Pluggable

(CFP)[1], is used to generate optical coherent signal. We have used CFP modules with baud rate of 32 Gbaud/s and quadrature phase shift keying (QPSK) modulation.

Note that in Figure 1, $\lambda_i$ denotes the wavelength of optical transmitter, where $\lambda_1 = 1552.52 nm$, $\lambda_2 = 1553.33 nm$, and $\lambda_3 = 1554.13 nm$. Furthermore, Array Waveguide Grating (AWG) WDM multiplexer (AWG-MUX) and de-multiplexer (AWG-DMUX) are used to multiplex and de-multiplex signals of three CFPs, respectively. It should be noted that, although in the single-channel scenario AWG-MUX and AWG-DMUX are not required, we used them to have the same optical attenuation in both arrangements. Moreover, we used four Erbium-doped fiber amplifiers (EDFAs), where $i$ th optical amplifier (OA) is denoted as $OA_i$. In all experiments, unless mentioned, the gain of $OA_1, OA_2, OA_3$, and $OA_4$ are set 18 dB, 25dB, 25dB, and 25dB, respectively.

We have used optical modulation analyzer (OMA) to monitor and measure characteristics of CFP output. In Fig. 2, a snapshot of the output of OMA is shown. In this figure, modulation constellation, I/Q eye pattern, and SNR, etc. are shown. We observe that both x- and y-polarizations are modulated based on QPSK modulation format, and input SNR is about 16.7 dB.

In order to evaluate non-linearity effects of SMF and NZDSF fibers, the transmit power of CFPs is swept from -22 dBm to -1 dBm (-3dBm) in single-channel (three-channel) scenario. Recorded results are presented in the next section.

## V. NUMERICAL RESULTS

### A. Analytical Results

In this section, we evaluate the results of optimization problems P2 and P4 solved by using CVX tool [14]. The considered fiber and other components parameters are listed in Table 1.

Table 1. Considered Parameters in Optimizations

| Parameter | Value |
| --- | --- |
| $\alpha$ | 0.2 $dB/km$ |
| $\beta_2$ | -21.45562 $ps^2/km$ |
| Fiber Optic Dispersion | 16.75 $ps/nm.km$ |
| Optical Frequency | 193 $THz$ |
| $\beta_3$ | 0 |
| $\gamma$ | 1.31 $(W.km)^{-1}$. |
| Span Length, $L_s$ | 120 $km$ |
| $n_{sp}$ | 1.77 |
| Modulation Format | PM-QPSK |
| Required SNR (all channels) | 8.45 dB |
| Baud Rate | 27.5 Gbaud/sec |
| Number of Spans | 40 |
| Number of Channels | 30 |

By solving the optimization problem P2 based on the parameters given in Table 1, we obtained optimum power of each channel in a 30-channel WDM system. Figure 3 shows the optimal power allocated for each channel and the results of flat power optimization ( $\mathbf{x} = x \times \mathbf{I}$ ). As can be seen, the central channel has higher transmitted power, because (as it was illustrated by DC-EGN and GN results) the central channel has high non-linear noise w.r.t. other channels, and to compensate its NLIN, higher power is allocated to this channel. We also observe that in case of variable power per channel index, the power of each channel is a concave function which is maximized in the central channel. It is also inferred that DC-EGN model needs 1.5 dBm of transmit power however, DC-GN model needs 1 dBm.

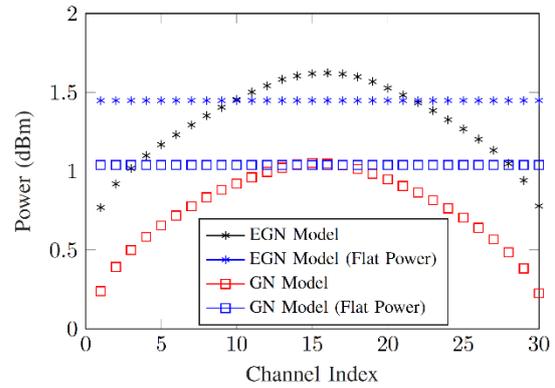

Figure 3. Results of power allocation by solving P2 based on either DC-EGN or GN models.

The power optimization based on the DC-EGN allocates higher power than the GN counterparts, which reveals that the DC-EGN based optimization can reach higher rate Figure 4 shows the average achievable rate per channel obtained by solving the rate maximization problem expressed in section III, where we observe that by using the DC-EGN model, higher rate per channel is achieved. This is due to the fact that the GN model overestimates NLIN and as a result it allocates lower power per channel which leads to lower SNR and achievable rate.

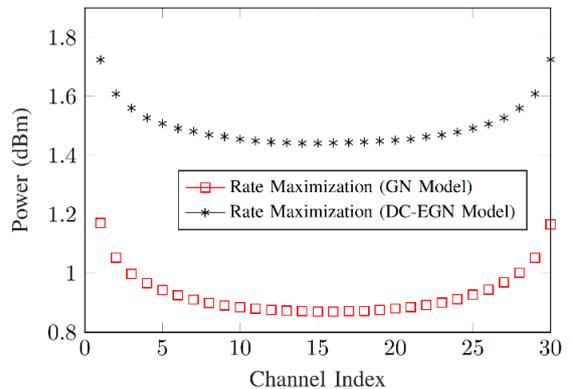

Figure 4. Results of power allocation by solving P4 based on both DC-EGN and GN models.

---
[1] Note that C stands for Latin *centum* (hundred), as CFPs have been designed to support data rate of 100Gbps.

Table 2. Number of Equality Constraints and Variables in Optimization Problems Based on DC-EGN and GN

| | 10-WDM | | 20-WDM | | 30-WDM | | 40-WDM | |
|---|---|---|---|---|---|---|---|---|
| | DC-EGN | GN | DC-EGN | GN | DC-EGN | GN | DC-EGN | GN |
| Number of equalities in optimization | 220 | 1360 | 840 | 10720 | 1860 | 36080 | 3280 | 85440 |
| Number of variables in optimization | 860 | 4040 | 3320 | 32080 | 7380 | 108120 | 13040 | 256160 |

To have a better understanding of the results of the DC-EGN model versus the GN one, in Figure 5 we show the achievable rates obtained by solving P4, where we observe that the DC-EGN based optimization achieves higher spectral efficiency (in terms of bit rate per hertz). For example, in case of 10 channels, the achieved rate is 12.95 bps/Hz while for GN-based model it is 12.78 bps/Hz.

In order to compare the complexity of optimization problems based on DC-EGN and GN models, we have recorded the number of equality constraints and variables of each optimizations, reported by CVX tool. In Table 2, the number of equality constraints and variables are shown for different number of WDM channels. As can be seen, the DC-EGN based optimizations have less variables and equality constraints which highlighting its lower computational complexity.

Figure 6 represents SNR versus channel index obtained by solving P4 for two scenarios in which one of them considers an equal power for all channels and the other one finds the power of each channel individually during the optimization. As can be seen, the optimization based on the DC-EGN model leads to higher SNR. For instance, in case of flat power for DC-EGN model, we reach 8.8 dB while for DC-GN model the SNR value is 8.1 dB; which shows 0.7 dB improvement in SNR.

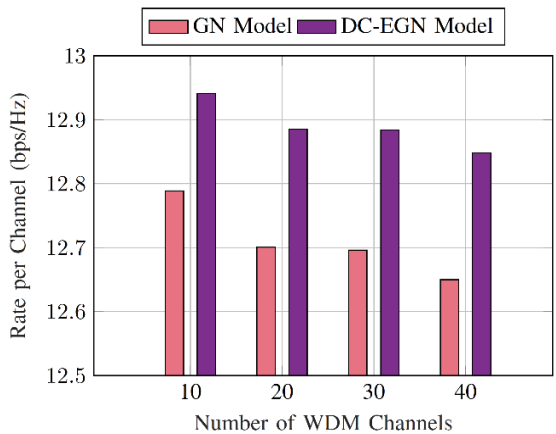

Figure 5. Comparison of Capacities Achieved by Solving P4 using DC-EGN and GN model

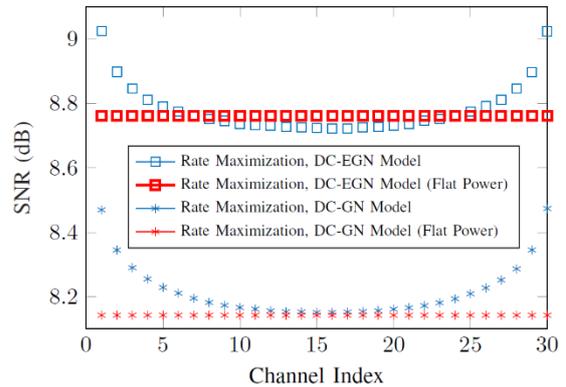

Figure 6. SNR Comparison of Solving P4 with DC-EGN and GN Models

*B. Simulation Results*

In what follows, we validate the results of optimizations by comparing them with the results of simulations. To this aim, we considered a 110 Gbps CUT system based on the Polarization Multiplexed Quadrature Phase Shift Keying (PM-QPSK) modulation format which has been simulated in the OptiSystem software. It is worth mentioning that in our experimental setup, we only had transceivers with PM-QPSK modulation format, thus to have consistent comparisons between simulation and experimental results, here we only implemented PM-QPSK. We considered a 5-channel WDM system and recorded bit-error-rate (BER) of the central channel (third channel). The baud rate of each channel is 27.5 Gbaud/sec and channel spacing is 100 GHz. In the simulation setup, the BER of the desired channel (third channel) is computed by counting error bits, which are determined by comparing transmitted and detected bits at the output of receiver.

By sweeping the number of spans $N_s$, we recorded BER and plotted it as a function of transmit power, as shown in Figure 7. It is clear that at lower transmit power the fiber is in linear regime; nonetheless, at higher transmit power the fiber non-linearity becomes dominant, and as a result, we have a point in which the BER achieves its minimum value which is occurred at an optimal power. In addition, with the growth of number of spans, $N_s$ the total NLIN increases, and as a result, BER is degraded.

To obtain the SNR of simulated WDM system, we considered the following theoretical BER formula for PM-QPSK [12]:

$$BER = \frac{1}{2}\text{erfc}\left(\sqrt{0.5 \times SNR}\right), \qquad (26)$$

where erfc(.) denotes error function. By using (26) we can find the SNR as follows:

$$SNR = 2 \times \text{erfc}^{-1}(2 \times BER), \quad (27)$$

where $\text{erfc}^{-1}(.)$ is the inverse function of $\text{erfc}(.)$. By using (27), the SNR of simulations are obtained by using recorded BERs.

In Figure 8, the estimated SNR of simulations is depicted. It is observed that by increasing $N_s$, SNR is degraded. In addition, the optimal power of third channel which maximizes the SNR is around 2 dBm. To compare the simulation results with output of the proposed optimization problems, we considered the same setting as the simulations, and allocated power of a 5-channel WDM system by solving problem P2.

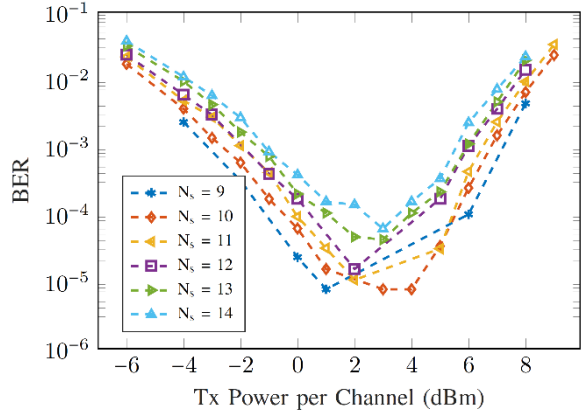

Figure 7. BER of third channel in the simulated 5-channel WDM System with 110 Gbps rate per channel

In next step, we obtained the results of power allocation with the objective of maximizing minimum SNR margin for the 5-channel WDM system. Interestingly, we observed that the optimal power of third channel for the case of flat power optimization is approximately 2.2 dBm which is in accordance with the simulation setting is close to the simulation results, which demonstrates the practical applicability of the proposed optimizations.

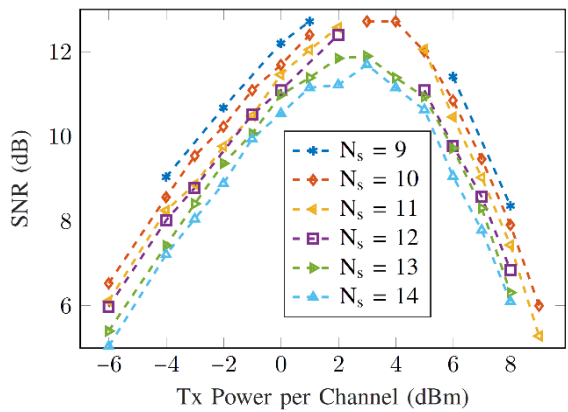

Figure 8. SNR of third channel in the simulated 5-channel WDM system with 110 Gbps rate per channel.

### C. Experimental Results

We have recorded SNR of coherent 100 Gbps system according to the two arrangements explained in Section IV. All results are obtained by recording reports of CFP which are available through the equipment management system (EMS) software. It is worth noting that we could not use OMA to measure SNR at the output of three spans. This is due to the fact that there is not any equalization module at the input of OMA to compensate fiber dispersion, thus, received signal at OMA cannot be analyzed. Consequently, we have used loop-back connection (i.e., received signal is loop backed to the RX of the CFP sending the input signal) as shown in Figure 1.

In Figure 9, measured SNRs for SMF in both arrangements (single-channel and three-channel) are shown versus the transmit power of CFP. We observe that with the increase of input power the non-linearity effects of SMF are dominated, which degrade SNR. As expected, three-channel scenario has a lower SNR at higher power, because XCI noise are appeared in this scenario. Note that in order to compare optimal power of experiments with analytical models, we have to plot results versus input power of the first span which is the output of $OA_1$.

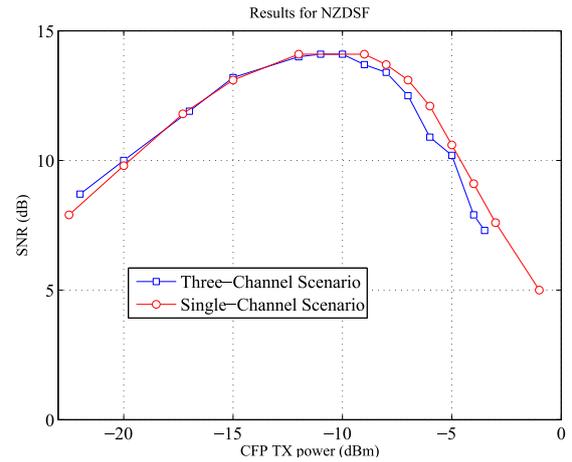

Figure 9. Recorded SNR for NZDSF versus TX power of CFP.

Table 3. Recorded Input/Output Powers at Optical Amplifier for the Case of Single-Channel SMF Experiment

| CFP TX $P_{out}$ (dBm) | $OA_1$ | | $OA_2$ | | $OA_3$ | | $OA_4$ | |
|---|---|---|---|---|---|---|---|---|
| | $P_{in}$ (dBm) | $P_{out}$ (dBm) | $P_{in}$ (dBm) | $P_{out}$ (dBm) | $P_{in}$ (dBm) | $P_{out}$ (dBm) | $P_{in}$ (dBm) | $P_{out}$ (dBm) |
| -22 | -23.5 | -5.6 | -28.7 | -3.7 | -31.3 | -6.3 | -25.6 | -0.6 |
| -20 | -21.2 | -3.2 | -26.9 | -1.9 | -30.4 | -5.5 | -25.3 | -0.3 |
| -18 | -19.6 | -1.6 | -25.6 | -0.6 | -29.7 | -4.7 | -25 | 0 |
| -15 | -16.3 | 1.6 | -22.6 | 2.3 | -27.6 | -2.7 | -24.1 | 0.8 |
| -12 | -13.3 | 4.6 | -19.7 | 5.2 | -25.3 | -0.3 | -22.8 | 2 |
| -9 | -10.3 | 7.6 | -16.8 | 8.1 | -22.7 | 2.2 | -21 | 3.9 |
| -8 | -9.3 | 8.6 | -15.8 | 9.1 | -21.8 | 3.1 | -20.3 | 4.6 |
| -7 | -8.3 | 9.6 | -14.8 | 10.1 | -20.9 | 4 | -19.6 | 5.5 |
| -6 | -7.3 | 10.6 | -13.8 | 11.1 | -19.9 | 5 | -18.8 | 6.1 |
| -5 | -6.3 | 11.6 | -12.8 | 12.1 | -19 | 5.9 | -18 | 6.9 |
| -4 | -5.3 | 12.6 | -11.8 | 13.1 | -18 | 6.9 | -17.1 | 7.8 |
| -3 | -4.3 | 13.6 | -10.8 | 14.1 | -17.1 | 7.8 | -16.2 | 8.7 |
| -2 | -3.3 | 14.6 | -9.8 | 15 | -16.1 | 8.8 | -15.3 | 9.5 |
| -1 | -2.3 | 15.6 | -8.8 | 16 | -15.1 | 9.8 | -14.4 | 10.5 |

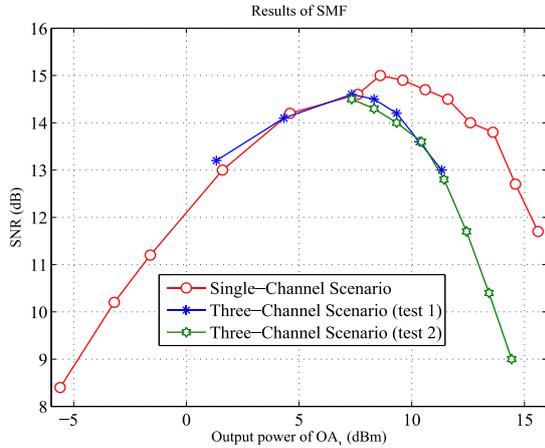

Figure 10. Recorded SNR for SMF versus out power of first optical amplifier (OA1).

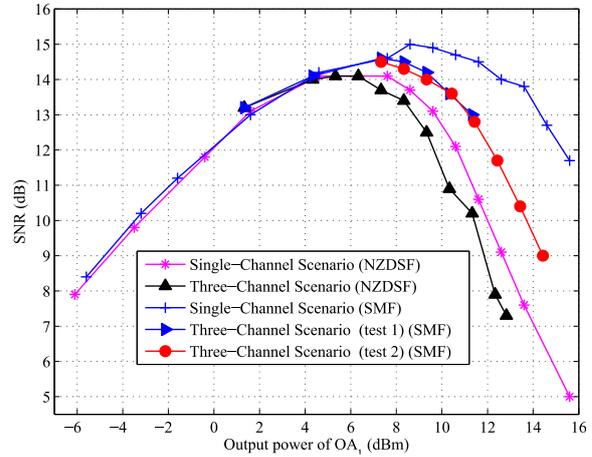

Figure 12. Comparing SNR of SMF and NZDSF versus output power of OA1.

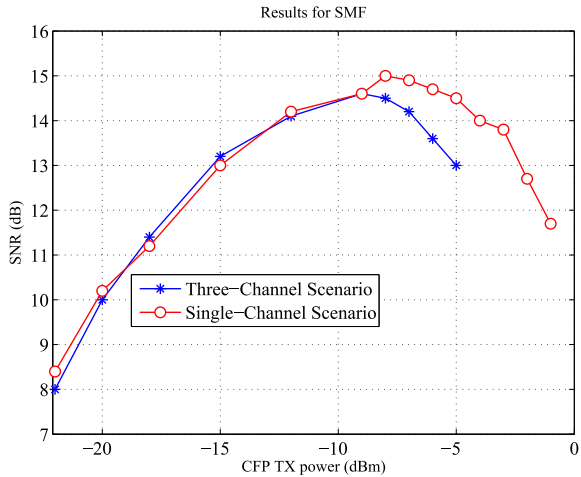

Figure 11. Recorded SNR for SMF versus TX power of CFP.

It is worth mentioning that in the three-channel scenario output power of CFPs were limited to -5dBm, because, for higher input power the total input power of $OA_1$ exceeds 0 dBm, which saturate optical amplifiers. Thus, in order to increase output power of $OA_1$, we have considered a new test case, referred to as test 2, in which the gain of $OA_1$ is increased to 25 dB. In Figure 10, we have plotted results of Figure 9, with respect to output power of $OA_1$ and added results of test 2. Note that in this figure test 1 means that the gain of $OA_1$ is

Table 4. Recorded Input/Output Powers at Optical Amplifier for the Case of Three-Channel SMF Experiment

| | $OA_1$ | | $OA_2$ | | $OA_3$ | | $OA_4$ | |
|---|---|---|---|---|---|---|---|---|
| CFP TX $P_{out}$ (dBm) | $P_{in}$ (dBm) | $P_{out}$ (dBm) | $P_{in}$ (dBm) | $P_{out}$ (dBm) | $P_{in}$ (dBm) | $P_{out}$ (dBm) | $P_{in}$ (dBm) | $P_{out}$ (dBm) |
| -22 | -13.4 | 4.5 | -19.9 | 5 | -25.2 | -0.2 | -23.4 | 1.5 |
| -20 | -13.1 | 4.8 | 19.5 | 5.4 | -25 | 0 | -23.2 | 1.7 |
| -18 | -11.1 | 6.8 | -17.6 | 7.3 | -23.2 | 1.7 | -22 | 2.9 |
| -15 | -11.9 | 6.1 | -18.4 | 6.5 | -23.9 | 1 | -22.5 | 2.4 |
| -12 | -8.8 | 9.1 | -15.5 | 9.4 | -21.3 | 3.7 | -20.5 | 4.4 |
| -9 | -5.8 | 12.1 | -12.4 | 12.5 | -18.4 | 6.5 | -18.1 | 6.8 |
| -8 | -4.8 | 13.1 | -11.4 | 13.5 | -17.4 | 7.4 | -16.6 | 7.7 |
| -7 | -3.8 | 14.1 | -10.4 | 14.5 | -16.5 | 8.3 | -15.7 | 9.2 |
| -6 | -2.8 | 15.1 | -9.4 | 15.5 | -15.5 | 9.4 | -15.4 | 9.5 |
| -5 | -1.8 | 16.1 | -8.6 | 16.3 | -14.6 | 10.3 | -14.5 | 10.4 |

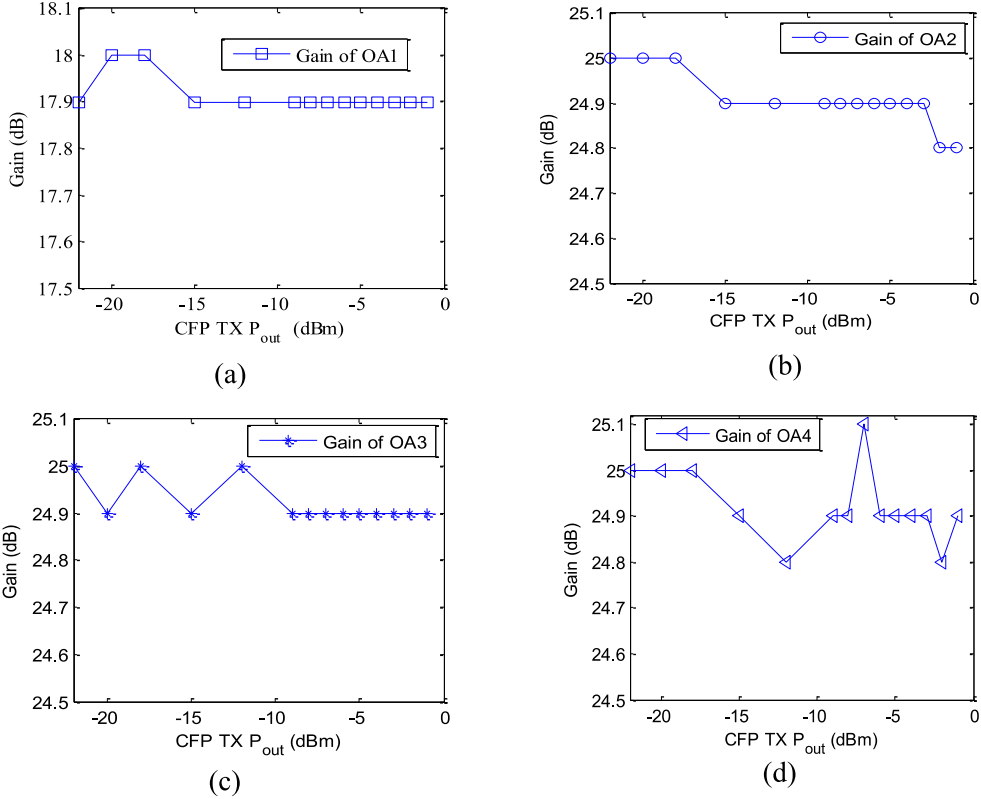

Figure 13. Measured gains of (a) OA1, (b) OA2, (c) OA3, and (d) OA4 for the single-channel and SMF experiments.

18 dB. We observe that by increasing the gain of $OA_1$ we could measure the non-linearity effect of SMF at higher powers. Results show that the optimum power in the single-channel and three-channel scenarios are +8.6 dBm and +7.3 dBm, respectively.

In Figure 11, SNRs of single-channel and three-channel scenarios for NZDSF are shown versus the transmit power of CFP. We observe that the results of NZDSF and SMF (as shown in Figure 9) have the same trend, unless in the NZDSF cases we have a lower SNR than the SMF scenarios. In order to have a better comparison, in Figure 12 we have shown the results of SMF and NZDSF versus the output power of $OA_1$ (input power of first span).

In Figure 12, we observe that the maximum SNR of single-channel scenario in SMF is 1 dB higher than NZDSF, and in the three-channel scenario this gap is reduced to 0.5 dB. The optimum powers in which we obtain maximum SNR in the single-channel and three-channel scenarios for SMF and NZDSF are +8.6 dBm, +7.3 dBm, +4.8 dBm, and +4.8 dBm, respectively. In addition, we observe that the SNR degradation with the increase of the number of channels in SMF is more than NZDSF.

In order to measure the power of intermediate spans, we have recorded input and output powers of each optical amplifiers which are presented in Table 3. In this table, output power is the power of amplified input signal and ASE noise. In Table 3, the results of single-channel scenario for SMF are reported. These results for the case of three-channel in SMF are presented in Table 4. In both tables we observe that the difference of output and input powers (which equals to the gain of optical amplifier) is not fixed. In order to validate the correctness of our results, we have plotted the differences of input and output powers of all optical amplifiers versus transmit power of CFP (input power of the system).

In Figure 13, gains of all optical amplifier based on the measured input and output powers are shown. Note that we have set the gain of $OA_1$ to 18 dB and the gains of $OA_2$, $OA_3$, and $OA_4$ to 25 dB, via EMS software. Results indicate that there is approximately $\pm 0.2$ dB fluctuation around the set values in the EMS. This fluctuations are acceptable, because, the amplification gain of optical amplifiers is slightly affected with the random behavior of ASE noise.

In Figure 14, the analytical (EGN model) and experimental results are compared for the three-channel scenario. We observe that there are large gaps between EGN and experimental results. This is due to the fact that there are other sources of noise in the experimental setup which are not included in the EGN model. By using the following relation we can improve the accuracy of the EGN model (inspired form Eq. (80) in [9]).

$$SNR_{EGN-TX_{noise}} = \frac{1}{\frac{1}{SNR_{EGN}} + \frac{1}{SNR_{input}}} \quad (28)$$

where $SNR_{EGN}$ is the SNR obtained from EGN model, and $SNR_{input}$ denotes the input SNR indicating imperfection of transmitter and other sources of noise. In Figure 14, the computed $SNR_{EGN-TX_{noise}}$ for $SNR_{input}$ = 16.7dB are shown. Note that we have measured the value of $SNR_{input}$ by using OMA as shown in

$$P_c^{NL}(e^{\mathbf{y}}) = e^{3y_c} D_1(c) + \sum_{n=1, n \neq c}^{N} [e^{y_c + 2y_n} D_2(c,n) \\ + e^{2y_c + y_n} D_3(c,n) + e^{3y_n} D_4(c,n)] \quad (19)$$

We observe that by considering (28), the gaps between analytical and experimental results are reduced. However, the closeness between analytical and experimental results is only for the pseudo-linear region (i.e., around the optimum power of EGN model), and for high-nonlinear region, the performance gap is increased. It is worth noting that this finding is in agreement with the results of [8], [9].

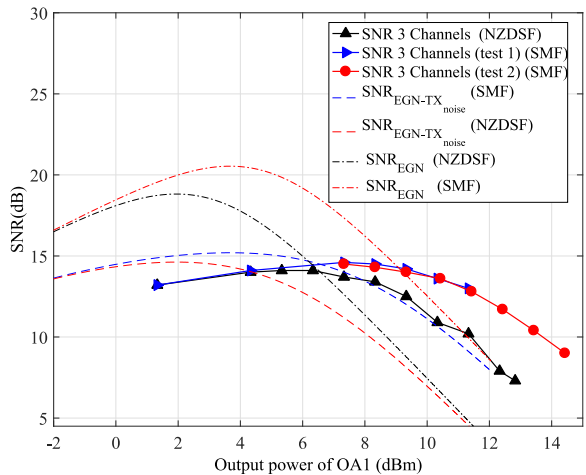

Figure 14. Comparing experimental and analytical results of the three-channel scenario.

## VI. CONCLUSION

In this paper, the problem of power allocation in optical coherent and uncompensated transmission system has been investigated. Two optimization problems have been introduced with the objective of maximizing minimum SNR margin and achievable rate, in which SNR of CUT is estimated based on the DC-EGN model. In the DC-EGN model, non-linear interferences namely, SCI and XCI on each channel are recorded in a lookup table. Then, by considering the index number of channels, the NLIN on the channel of interest is computed and accordingly SNR is obtained by considering the transmission power and optical amplifier noise. We evaluated the proposed optimization problems and compared them with the existing alternatives which are based on the discretized GN model. We observed the superiority of the proposed optimization problems in terms of computational complexity, SNR, and achievable rate. In addition, we validated our results with simulations, whereby we observed a close agreement between optimization results and simulations. In addition, we presented results of experimental setups in which we measured SNR of commercial single-channel and three-channel 100 Gbps coherent systems. We investigated the non-linear effects of SMF and NZDSF, and compared experimental results and EGN model. We showed that there are large gaps between EGN model and experimental results which can be reduced by considering other sources of noise such as imperfection of transmitter, thermal noise, and shot noise.

ACKNOWLEDGMENT

This work was supported in part by Iran Telecommunication Research Center, Tehran, Iran. A part of this paper has been presented in 9th International Symposium on Telecommunications (IST'2018), Tehran, Iran.

REFERENCES

[1] Kazuro Kikuchi. Fundamentals of coherent optical fiber communications. *Journal of Lightwave Technology*, 34(1):157–179, 2016.

[2] Ori Gerstel, Masahiko Jinno, Andrew Lord, and SJ Ben Yoo. Elastic optical networking: A new dawn for the optical layer? *IEEE Communications Magazine*, 50(2), 2012.

[3] H. Beyranvand, M. Maier, and J. A. Salehi. An Analytical Framework for the Performance Evaluation of Node- and Network-Wise Operation Scenarios in Elastic Optical Networks. *IEEE Transactions on Communications*, 62(5):1621–1633, May 2014.

[4] Mehrdad Moharrami, Ahmad Fallahpour, Hamzeh Beyranvand, and Jawad A Salehi. Resource Allocation and Multicast Routing in Elastic Optical Networks. *IEEE Transactions on Communications*, 65(5):2101–2113, 2017.

[5] E. Ehsani Moghaddam, H. Beyranvand, and J. A. Salehi. Routing, Spectrum and Modulation Level Assignment, and Scheduling in Survivable Elastic Optical Networks Supporting Multi-Class Traffic. *Journal of Lightwave Technology*, 36(23):5451–5461, December 2018.

[6] Hamzeh Beyranvand and Jawad A Salehi. A quality-of-transmission aware dynamic routing and spectrum assignment scheme for future elastic optical networks. *Journal of Lightwave Technology*, 31(18):3043–3054, 2013.

[7] Kim Roberts, Qunbi Zhuge, Inder Monga, Sebastien Gareau, and Charles Laperle. Beyond 100 Gb/s: capacity, flexibility,

and network optimization. *IEEE/OSA Journal of Optical Communications and Networking*, 9(4):C12–C23, 2017.

[8] Pierluigi Poggiolini, G Bosco, A Carena, V Curri, Y Jiang, and F Forghieri. The GN-model of fiber non-linear propagation and its applications. *Journal of lightwave technology*, 32(4):694–721, 2014.

[9] Pierluigi Poggiolini, Gabriella Bosco, Andrea Carena, Vittorio Curri, Yanchao Jiang, and Fabrizio Forghieri. A detailed analytical derivation of the GN model of non-linear interference in coherent optical transmission systems. *arXiv preprint arXiv:1209.0394*, 2012.

[10] Ronen Dar, Meir Feder, Antonio Mecozzi, and Mark Shtaif. Properties of nonlinear noise in long, dispersion-uncompensated fiber links. *Optics Express*, 21(22):25685–25699, 2013.

[11] Andrea Carena, Gabriella Bosco, Vittorio Curri, Yanchao Jiang, Pierluigi Poggiolini, and Fabrizio Forghieri. EGN model of non-linear fiber propagation. *Optics Express*, 22(13):16335–16362, 2014.

[12] Ian Roberts, Joseph M Kahn, and David Boertjes. Convex channel power optimization in nonlinear WDM systems using Gaussian noise model. *Journal of Lightwave Technology*, 34(13):3212–3222, 2016.

[13] Shiva Kumar and M Jamal Deen. *Fiber optic communications: fundamentals and applications*. John Wiley & Sons, 2014.

[14] M Grant and S Boyd. CVX: Matlab Software for Disciplined Convex Programming, version 2.0. *Recent Advances in Learning and Control*, pages 95–110, 2008.

[15] ITU-T. G. 709: Interfaces for the Optical Transport Network (OTN), 2003.

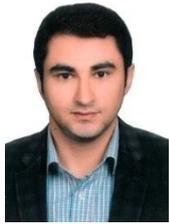

**Ramin Hashemi** received the B.Sc. and M.Sc. degrees (with honors) from Amirkabir University of Technology (Tehran Polytechnic), Tehran, Iran, in 2016 and 2018, respectively. Since September 2018, he has been with Microwave/Millimeter-Wave & Wireless Communication Research Lab. (MMWCL) as a graduate research assistant at the Amirkabir University of Technology. His research interests include fiber optic communications, resource allocation in fiber-wireless networks, MIMO wireless communications and beyond 5G.

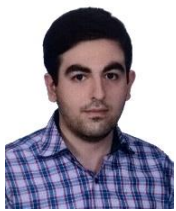

**Mehdi Habibi** received the B.Sc. and M.Sc. degrees from Amirkabir University of Technology (AUT), Tehran, in 2018 and 2016, respectively. His research interests are optical communcations and networking, elastic optical networks and network optimization.

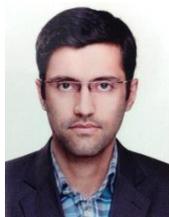

**Hamzeh Beyranvand** is an assistant professor at the Electrical Engineering Department, Amirkabir University of Technology, Tehran, Iran. He received his PhD in Electrical Engineering from the Sharif University of Technology (SUT) in 2014. He received the 14th Iran National Khwarizmi Youth Award Dec. 2012 (Ranked 2 in Applied Research), and the best PhD thesis award from IEEE Iran Section, 2015. His research interests are: fiber optic communications and networking, optical wireless communications, broadband access networks, and network planning and optimization.

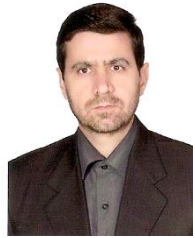

**Mahdi Hashemi** received his M.Sc. in Physics from Sharif University of Technology in 1994. He is currently working as a researcher in ICT Research Institute (Iran Telecommunication Research Center). He has more than 20 years of working experience in Optical Communications and the related industry. His current research interests are in the areas of optical and quantum communications.

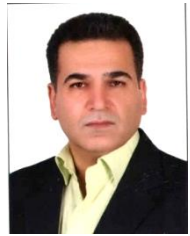

**Ali Emami** received his M.Sc. in Electronic from Azad University South Tehran Branch in 2003. He is currently working as a Faculty member in ITRC (Iran Telecommunication Research Center). He has more than 16 years of working experience in Optical Communications and the related industry. His current research interests are in the areas of optical Transmission.

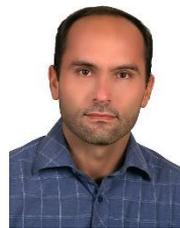

**Davood Ranjbar Rafi** received his M.Sc. in ICT Communication from Amirkabir University of Iran in 2005. He is currently working as a researcher in ICT Research Institute (Iran Telecommunication Research Center). He has more than 24 years of working experience in Optical Communications and the related industry. His current research interests are in the areas of optical access and Transmission communications.